\documentclass[twocolumn,prl,showpacs]{revtex4}

\usepackage{graphicx}
\usepackage{dcolumn}
\usepackage{bm}
\usepackage{amsxtra}
\usepackage{latexsym}

\newcommand{\ket}[1]{\mbox{\ensuremath{ | #1 \rangle }}}                           

\newcommand{\inprod}[2]{\mbox{\ensuremath{ \langle #1 | #2 \rangle}}}

\newcommand{\cop}[1]{\mbox{\ensuremath{ a_{#1}^{\dagger} }}}
\newcommand{\aop}[1]{\mbox{\ensuremath{ a_{#1}^{\phantom{\dagger} }}}}
\newcommand{\op}{\mbox{\ensuremath{P_{\infty}  }}}
\newcommand{\dashline}{\mbox{\protect\rule[2pt]{0.1cm}{0.1pt} \hspace{-4pt} 
    \protect\rule[2pt]{0.1cm}{0.1pt} \hspace{-4pt} \protect\rule[2pt]{0.1cm}{0.1pt}}}
\newcommand{\ddashline}{\mbox{$\cdot \cdot$ \hspace{-3pt}
    \protect\rule[2.1pt]{0.175cm}{0.6pt} \hspace{-3pt}
    $\cdot \cdot$}}

\begin{document}

\preprint{The contact process in heterogeneous and weakly-disordered systems }  
  
\title{The contact process in heterogeneous and weakly-disordered systems}  
  
\author{C.~J.~ Neugebauer}  
 \email{cjn24@cam.ac.uk}  
\affiliation{Department of Chemistry, University of Cambridge, Cambridge, UK}

\author{S.~V.~ Fallert}  
\affiliation{Department of Chemistry, University of Cambridge,  
             Cambridge, UK}

\author{S.~N.~Taraskin}  
\affiliation{St. Catharine's College and Department of Chemistry, University of Cambridge,  
             Cambridge, UK}  

\date{\today}
  
\begin{abstract} 
The critical behavior of the contact process (CP) in heterogeneous periodic and weakly-disordered
environments is investigated using the supercritical series expansion
and Monte Carlo (MC) simulations.
Phase-separation lines and critical exponents $\beta$ (from series expansion) and
$\eta$ (from MC simulations) are calculated.
A general analytical expression for the locus of critical points is suggested
for the weak-disorder limit and confirmed
by the series expansion analysis and the MC simulations.
Our results for the critical exponents show that the CP in heterogeneous
environments remains in the directed percolation (DP) universality class, 
while for environments with quenched disorder, the data are compatible with
the scenario of continuously changing critical exponents.
\end{abstract}  
  

  
\maketitle  
  

Phase transitions in non-equilibrium statistical mechanics have long been a 
field of interest, and
universality classes similar to those in equilibrium 
have been identified, with the DP class being one of
the most prominent ones. 
The CP \cite{harris_74}, 
a susceptible-infected-susceptible  model for the spread of epidemics, 
belongs to this universality class and has become 
 one of its archetypical models. 
Recent years have seen much activity sparked by the
question whether the DP class is robust with respect 
to the introduction of quenched spatial disorder or not. 
This has
been seen as an important question both from a fundamental and  experimental 
viewpoint, as it has been suggested that the
lack of experimental observations of the DP critical behaviour might be a
result of the presence of disorder in real-world systems~\cite{hinrichsen_00}.
 
One of the foremost arguments that disorder changes the critical behaviour of
the CP is that it violates the Harris criterion \cite{harris_74_2,chayes_86} 
for all dimensions $d < 4$. 
This criterion
states that a critical
point is stable with respect to disorder if $d \nu_{\bot} > 2$ where 
$\nu_{\bot}$ is the
critical exponent associated with the spatial correlation length.
So far, all studies carried out on the disordered CP have provided supporting
evidence for a change in universality with the introduction of
disorder 
\cite{noest_85,dickman_96,dickman_98,janssen_97,hooyberghs_03,hooyberghs_04,vojta_05}. 
However, it is not entirely clear how the critical exponents change with disorder.
In the strong-disorder limit,  
Hooybhergs et al.~\cite{hooyberghs_03,hooyberghs_04} 
have demonstrated that the CP changes to the universality class 
of the random transverse-field Ising model with activated scaling 
characterized by known scaling exponents. 
Recent MC simulations \cite{vojta_05} 
suggest that the activated scaling holds for an arbitrary degree of disorder 
meaning that an introduction of even weak disorder forces the abrupt 
change of critical exponents from known values for the homogeneous CP 
to those of the infinite-randomness fixed point (IRFP).  
This contradicts the findings of other authors 
\cite{hooyberghs_04,dickman_98}, 
who showed, both using MC simulations and 
density-matrix renormalization-group (RG) analysis, 
that there is an intermediate disorder regime with continuously varying
exponents. 
Unconventional critical behaviour produced by quenched randomness 
is supported by 
a field-theoretical analysis
~\cite{janssen_97} 
in which only runaway solutions in the RG equations were found.

The subject of this paper is to investigate the $1d$ CP in 
heterogeneous periodic systems (e.g.~a regular binary chain) 
and in systems with weak disorder (e.g.~a binary chain with 
randomly placed species characterized by parameters close in value). 
In order to achieve this goal, we employ 
the supercritical series expansions \cite{dickman_91,jensen_93} 
and MC simulations. 
We also suggest a simple analytical expression for the 
locus of critical points in the rate-space phase diagram which is 
in very good agreement with series expansion and MC simulation 
results both for heterogeneous periodic and weakly-disordered systems. 
Our main findings demonstrate that the CP in heterogeneous periodic 
systems belongs to the DP universality class with the scaling 
exponents coinciding with those for homogeneous case.      
For weakly-disordered systems, we can state that the introduction 
of disorder does not force the exponents to change to the 
values of the IRFP but rather 
causes their continuous change with disorder strength. 

The CP is usually defined on a hypercubic lattice of nodes 
which can be either empty (susceptible) or occupied (infected). 
The infection occurs via contacts between $Z$ nearest nodes $i$ and 
$j$ with the rate $\lambda_{ij}/Z$. 
An infected node $i$ can recover to susceptible one with the rate $\mu_i$. 
The time scale is defined by setting all $\lambda_{ij}=1$ 
(for simplicity, there is no disorder in transmission rates) and, 
for concreteness, we consider only binary systems with two types 
of nodes, $A$ and $B$, characterized by the recovery rates, 
$\mu_{A}$ and $\mu_B$, respectively, 
which are distributed according to bimodal 
distribution in the disordered system, $\rho(\mu_i)= (1-q)\delta(\mu_i-\mu_A) 
+ q \delta(\mu_i-\mu_B)$, with $q$ being the concentration of nodes $B$.

In the homogeneous case ($q=0$), the CP 
undergoes a non-equilibrium phase transition between 
active and absorbing states \cite{liggett_85} 
if the recovery rates become greater than a critical value, 
$\mu > \mu_c\simeq 0.303228$ \cite{jensen_93}. 
At criticality, the number of infected sites, $N_{\text{inf}}(t)$, scales with time 
as, $N_{\text{inf}}(t)\propto t^{\eta}$, with $\eta=0.313686$ 
\cite{jensen_96}. 
Close to criticality,    
the survival probability, $P_\infty$, and the time and space 
correlation lengths,  $\xi_{||}$ and $\xi_{\bot}$, respectively,   
also exhibit typical critical behaviour, 
$P_{\infty} \propto  \Delta^{\beta}$, 
$ \xi_{||} \propto |\Delta|^{-\nu_{||}} $ and 
$\xi_{\bot} \propto |\Delta|^{-\nu_{\bot}}$ where 
$\Delta = \mu_c - \mu$ and the universal exponents are 
$\beta\simeq 0.2769$ \cite{jensen_93}, 
$\nu_{||} = 1.733825(25)$ and $\nu_{\bot} = 1.096844(14)$ \cite{jensen_96}. 

In the heterogeneous and disordered systems, a similar transition 
occurs \cite{bramson_91}. 
Below, we address two questions: (i) how  
heterogeneity and disorder influence  
the universal properties (namely, scaling exponents) at criticality 
and (ii) what  the locus of critical points is in the rate space 
of the CP in such systems. 
We start with simple arguments about a possible way of evaluating 
an analytical expression for the critical line separating 
the active and absorbing states. 

Let us consider a system of $N$ ($N\to\infty$) nodes characterized 
by random recovery rates, $\tilde{\mu}_i = \mu_i/\mu_c$. 
The locus of critical points in the space of recovery rates can be 
written in a general form as a solution of the 
following equation, $F(M_i)=0$ (for $i = 1,2,\ldots,N$), 
where $M_i = \ln\tilde{\mu}_i$.  
The function $F(M_i)$ is invariant under the exchange of any two
arguments and it obeys the property $F(0)=0$ for the homogeneous case.  
Assuming that this function is analytic around the homogeneous critical 
point we can expand it in a Taylor series around this point, 
$ F'(0)\sum_i M_i + O(M_i^2)=0$, 
where we have used the symmetry of the function 
$F$ leading to all first derivatives being equal at the stationary point. 
In fact, this expansion is equivalent to the expansion in the moments 
$u_n = E[M^n]$, 
where $E[\cdot]$ denotes the expectation value. 
Leaving only the first order in the above expansion we end up 
with the following approximate equation for the locus of critical points, 
\begin{equation}
E[\ln\tilde{\mu}]\simeq 0~,
\label{e1}
\end{equation}
which is valid around the homogeneous critical point.  
The choice of arguments $M_i$ of function $F$ 
is motivated by a similar choice for the RG analysis 
of both the CP and random transverse-field Ising model 
\cite{fisher_95,hooyberghs_03}. 
Below, we demonstrate that Eq.~(\ref{e1}) indeed describes well the 
phase-separation lines in all studied heterogeneous and disordered systems. 

In order to support Eq.~(\ref{e1}) for the locus of critical points and also 
to investigate the scaling behaviour at criticality, we have 
used the perturbative supercritical series expansion 
\cite{dickman_91,jensen_93} for the survival probability, 
$P_{\infty}$.
Following the formalism developed in Refs.~\cite{doi_76,grassberger_79,peliti_85}, 
the state of the
system is described by the state vector 
$\ket{P(t)}=\sum_{\{\sigma\}}P (\{\sigma\},t)\,\ket{\{\sigma\}}$, whose time evolution is
governed by the master equation $\partial_t \ket{P(t)} = \hat{L}
\ket{P(t)}$. 
Here 
$\hat{L}$ is the generator of the Markov process that contains the transition rates
between the different microstates of the system, $\ket{\{\sigma\}}$.
As usual, $\hat{L}=\mu\,\hat{W} + \hat{V}$, is split into a part that 
destroys particles, $\mu \hat{W}$, and a part that creates particles, $\hat{V}$, 
which in the systems under consideration take the forms 
\begin{eqnarray}
\mu\,\hat{W} &=& \sum_i \mu_i\, (1-\cop{i})\,\aop{i} \label{eq:def_W}\\
\hat{V} &=& \sum_i \frac{1}{2}\,(1-\aop{i})\,\cop{i}\,
(\cop{i-1}\aop{i-1}+\cop{i+1}\aop{i+1})~, 
 \label{eq:def_V}
\end{eqnarray}
where $\cop{i}$ and $\aop{i}$ are hard-core bosonic creation and annihilation
operators, respectively.
The Laplace transform of $\ket{P(t)}$,
$\ket{\tilde{P}(s)}=(s-\mu\,\hat{W}-\hat{V})^{-1} \ket{P\,(0)}$, 
is then expanded in $\mu_i$, 
yielding the supercritical expansion for the
survival probability 
$P_{\infty}(\mu_A, \mu_B)=\lim_{s \rightarrow 0} (1 -
  s\,\inprod{0}{\tilde{P}\,(s)})$, where $\ket{0}$ is the absorbing state.
For the analysis of this multi-variable survival probability 
(cf. Ref.~\cite{dantas_05}),
we employ a numerical scheme similar to the 
\emph{Nested Pad\'e Approximation} \cite{guillaume_97,guillaume_00}. 
In order to investigate the critical behaviour, we 
consider the meromorphic function 
$\partial_{\mu_A} \ln \op(\mu_A, \delta)$ (with $\delta=\mu_B-\mu_A$),
whose first pole on the positive real axis is the critical point 
of the model and the residue at that pole is the critical exponent $\beta$. 
To improve estimates of these poles from the finite series expansion, 
the following multivariable rational-approximant scheme is used:
for a given expansion of $\op$ in two variables, 
$\mu_A$ and $\delta$, up to an even (odd) order
$N$,  the Pad\'e approximants $[n,n]$ ($[n,n+1]$) 
in $\delta$ of the coefficients of the
terms $\mu_A^{N-1-2n}$ ($\mu_A^{N-2n}$) in the series 
$\partial_{\mu_A} \ln \op(\mu_A, \delta)$ are formed,  
followed by the construction of the Pad\'e approximant 
[$N/2-1$, $N/2$] ([$(N-1)/2$, $(N-1)/2$]) in $\mu_A$.  
In order to estimate the stability of the poles and residues found, 
several Pad\'e approximants in $\mu_A$ (e.g.~the approximants from 
[$N/2-1$, $N/2$] down to [$N/2-2$, $N/2-1$] for even
orders of $N$) were constructed and averaged over.

\begin{figure}
\begin{center}
\scalebox{0.32}{\includegraphics[angle=0]{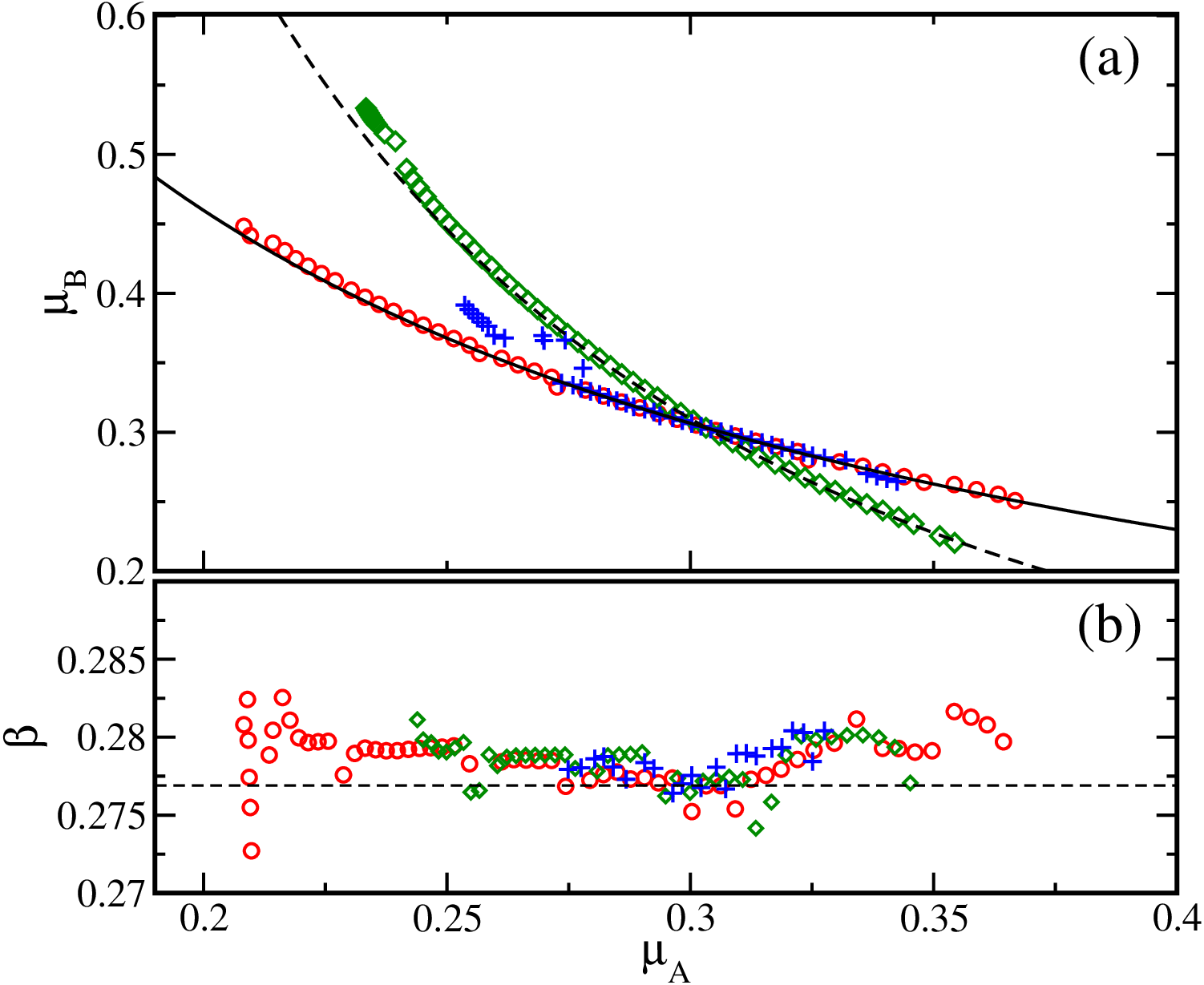}}
\end{center}
\caption{(Color online) Periodic 
$1d$  
lattices, $AB$ ($\circ$), $AAB$
  ($\diamond$), and $AABB$ ($+$): (a) critical points obtained by series 
  expansions in comparison with 
analytical prediction 
for the critical
  line $\mu_c = (\mu_A^{1-q}\,\mu_B^{q})$, $q = 1/2$ 
  (\protect\rule[2pt]{0.3cm}{0.1pt}) and $q = 1/3$ 
  (\dashline)
  for $\mu_c=0.303228$ \cite{jensen_93}; 
  (b) critical exponent $\beta$ in comparison with series expansion value
  $\beta = 0.2769$ (\dashline) 
  \cite{jensen_93} for the homogeneous case.} 
\label{fig:crystal_phase_diag}
\end{figure}

\begin{figure}
\begin{center}
\scalebox{0.32}{\includegraphics[angle=0]{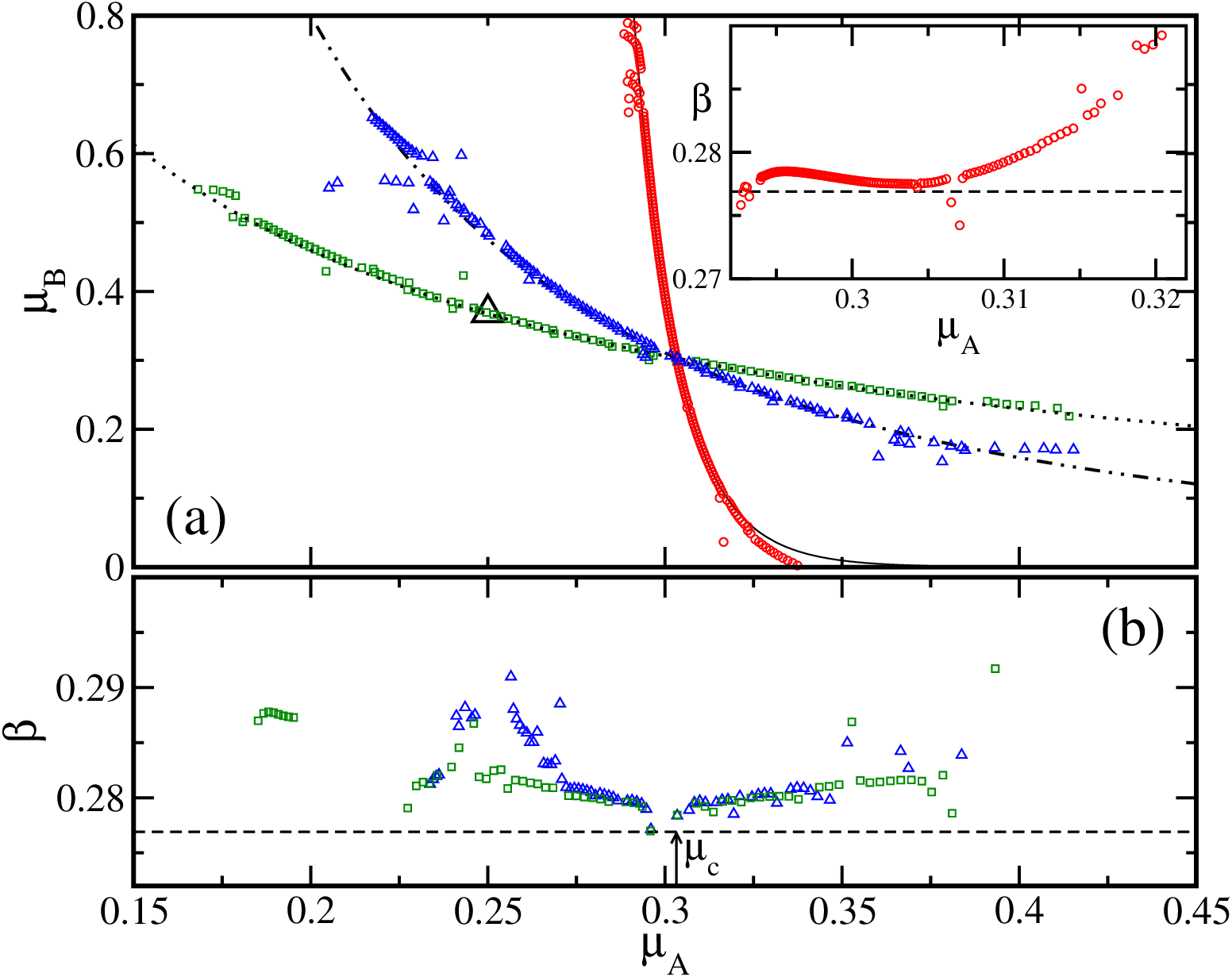}}
\end{center}
\caption{(Color online) The phase diagram (a) and scaling exponent
  $\beta(\mu_A)$, (b) and inset of (a), for disordered lattices with various 
  degrees of disorder: $q = 0.04$ ($\circ$),                                 
  $q = 0.3$ ($\triangle$) and $q = 0.5$ ($\Box$)
  obtained by series expansion. 
  The lines represent the theoretical prediction  according to Eq.~(\ref{e1}),   
  $\mu_c = (\mu_A^{1-q}\,\mu_B^q)$, for  
  $q = 0.04$ 
  (\protect\rule[2pt]{0.3cm}{0.1pt}), 
  $q=0.3$ (\ddashline) and $q=0.5$ ($\cdots$) 
  with $\mu_c=0.303228$ \cite{jensen_93}. 
  The triangle
  in (a) marks the region for which the MC
  simulations shown in  Fig.~\ref{fig:MC_bdcp} have been run. 
  The dashed lines in (b) and the inset of (a) show the value of $\beta$ for the homogeneous
  case, 
  $\beta_c = 0.2769$ (\dashline) \cite{jensen_93}. The arrow in (b) marks the
  homogeneous critical point.}  
\label{fig:bdcp_phase_diag}
\end{figure}

Using this scheme and starting from a single seed in the series expansions 
up to order $N=24$, 
we calculated the locus of critical points and the
critical exponents $\beta$ for three
heterogeneous lattices, $AB$, $AAB$, and $AABB$, and for disordered systems
whose recovery
rates 
 are drawn from the bimodal distribution mentioned above 
(see Figs.~\ref{fig:crystal_phase_diag} and ~\ref{fig:bdcp_phase_diag}). 
Fig.~\ref{fig:crystal_phase_diag}(a) demonstrates  
that around the homogeneous critical point ($\mu_c$, $\mu_c$),  
the phase-separation lines between the active and absorbing state are 
indeed very well 
described 
by Eq.~(\ref{e1}). 
This is also confirmed by 
single-seed 
MC simulations based on 
the random-sequential algorithm~\cite{marro_99} 
(see Table~\ref{tab:table1}) - the deviations of the MC results from 
predictions of Eq.~(\ref{e1}) and series expansion data for critical line 
are less than $1\%$. 
The critical values of $\mu_B$ for fixed values of $\mu_A$ were found by
analysis of the power-law behavior of $N_{\text{inf}}(t)$ with averaging over 
$10^6$ MC runs. 
Fig.~\ref{fig:crystal_phase_diag}(b) and Tab.~\ref{tab:table1} also show that the critical 
exponents $\beta$ and $\eta$ practically do not change from the 
values for the
homogeneous CP, thus confirming that the CP in the heterogeneous lattices
belongs to the same universality class, DP, as the homogeneous one.  
The systematic deviations of the calculated critical rates from the theoretical prediction
increases with the distance from the homogeneous point thus reflecting the
restricted range of applicability of Eq.~(\ref{e1}). 
Some irregular fluctuations both in $\mu_B$ and $\beta$ are probably due to
poor convergence of the series expansions.

\begin{table}
\caption{\label{tab:table1} 
The critical values of $\mu_B$ obtained in MC simulations (second column) and 
calculated according to Eq.~(\ref{e1}) (third column) together with 
the MC critical exponents (fourth column) for fixed values of $\mu_A$ 
(first column) in the heterogeneous systems $AB$ and $AAB$. 
}
\begin{ruledtabular}
  \begin{tabular}{ccccc}
           & $\mu_A$ & $\mu_B$ (MC)        & $\mu_B$ (pred.) & $\eta$ \\
    \hline
    $AB$   & 0.2750  & 0.3344 $\pm$ 0.0001 & 0.3343          & 0.313 $\pm$ 0.006\\
           & 0.2500  & 0.3681 $\pm$ 0.0001 & 0.3678          & 0.313 $\pm$ 0.003\\
           & 0.2250  & 0.4094 $\pm$ 0.0001 & 0.4087          & 0.313 $\pm$ 0.003\\
    $AAB$  & 0.2750  & 0.3689 $\pm$ 0.0001 & 0.3687          & 0.313 $\pm$ 0.002\\
           & 0.2500  & 0.4475 $\pm$ 0.0001 & 0.4461          & 0.314 $\pm$ 0.001\\
           & 0.2250  & 0.5546 $\pm$ 0.0001 & 0.5507          & 0.314 $\pm$ 0.001\\
\end{tabular}
\end{ruledtabular}
\end{table}
  
A similar series expansion analysis has been performed for disordered systems 
with a configurational averaging  of the
survival probability, $\langle \op \rangle$. 
We were able to perform the complete numerical averaging over all $2^{2 N - 1}$  configurations 
for $N\le 12$. 
The configurational averaging for series expansions to higher orders, $N = 19$, 
has been done approximately. 
In the calculation of $\langle \op \rangle= \sum_{n=0}^N \sum_{m=0}^n \langle c_{nm} \rangle
\mu_B^m \mu_A^{n-m}$, at each term of order $M \le N$, we only included realizations
with number $i \leq i_{M}$ of ``impurity''-sites $B$: e.g.~ we have chosen
$i_{N-n} = n + 2$ for series expansions up to order $N=19$, so that for $M =
18$ only realizations with up to three impurities
contributed to $\langle \op \rangle$.
Dropping these disorder realizations from the configurational average
incurs less error the smaller the impurity concentration $q$ 
is.
Assuming that the coefficients $c_{Mm} \sim c_M$ (with $c_M$ being the
coefficient of the same order in the homogeneous case),  
all the realizations with $i$
impurities contribute a term ${2 M - 1 \choose i} (1-q)^{2 M-1-i} q^i$ to the
configurational average, $\langle c_{Mm}\rangle$.   
Then, it can easily be shown that for $q=q_{\text{max}} = 0.04$ all terms with $i >
i_N = 2$ are smaller than the terms with $i \leq i_N$. 
The validity of this approximation has been confirmed by 
testing it against the exact results for $N\le 12$.

The results for fully and partially averaged survival
probabilities in disordered systems
are shown in Fig.~\ref{fig:bdcp_phase_diag}. 
The phase-separation lines have been obtained for arbitrary impurity
concentration for the fully averaged $\op$
expanded up to order $N=12$ and two of them for $q=0.5$ 
(squares and dotted line)
and $q=0.3$ 
(triangles and dot-dashed line)
 are displayed in Fig.~\ref{fig:bdcp_phase_diag}(a).  
High-order series expansions ($N = 19$) have been calculated only 
for low impurity concentrations, $q \le q_{\text{max}}=0.04$ 
(see the circles and  solid
line in Fig.~\ref{fig:bdcp_phase_diag}(a)). 
Again, the poles of 
$\partial_{\mu_A} \ln \op(\mu_A, \delta)$ agree very well with the theoretical
prediction given by  Eq.~(\ref{e1}).

The residues of the poles (exponents $\beta$) for different points on 
the critical line are shown in Fig.~\ref{fig:bdcp_phase_diag}(b) 
and in the inset in Fig.~\ref{fig:bdcp_phase_diag}(a). 
The value of $\beta$ reaches a minimum, $\beta_{\text{min}}$, 
located approximately around the homogeneous critical point $(\mu_c,\mu_c)$ 
with $\beta_{\text{min}}$ being rather close to the value of the homogeneous critical 
exponent, $\beta_c$, 
with $|\beta_{\text{min}}-\beta_c|/\beta_c \simeq 0.5 \%$ for $N=12$ and 
$0.21 \%$ for $N=19$ (see the inset in Fig.~\ref{fig:bdcp_phase_diag}(a)) 
thus confirming that the value of the exponent 
is much more sensitive to $N$ than the critical rates. 
Away from the critical point, the value of $\beta$, first, monotonically 
increases and then starts to fluctuate due 
to a high sensitivity to the value of
$\mu_B$, the estimates of which loose precision due to poor convergence of the
series in this range.  
The results for $\beta$ are in reasonable agreement with findings in 
Refs.~\cite{dickman_96,dickman_98,hooyberghs_04} 
where continuously varying critical exponents were seen in MC
simulations and density-matrix renormalization-group studies of the random CP. 
Unfortunately, the errors in the exponents that are shown in 
Fig.~\ref{fig:bdcp_phase_diag}(b) and the inset of Fig.~\ref{fig:bdcp_phase_diag}(a) 
are at least of the order of $|\beta(\mu_c)-\beta_c|$ 
and the monotonic growth of the exponents found in the series expansions 
can be questioned.  
However, our results are certainly not consistent with the scenario presented
in \cite{vojta_05} 
according to which the weakly-disordered CP belongs to
the same universality class as random transverse-field Ising model 
with $\beta = 0.38197$.

\begin{figure}
  \begin{center}
    \scalebox{0.32}{\includegraphics[angle=0]{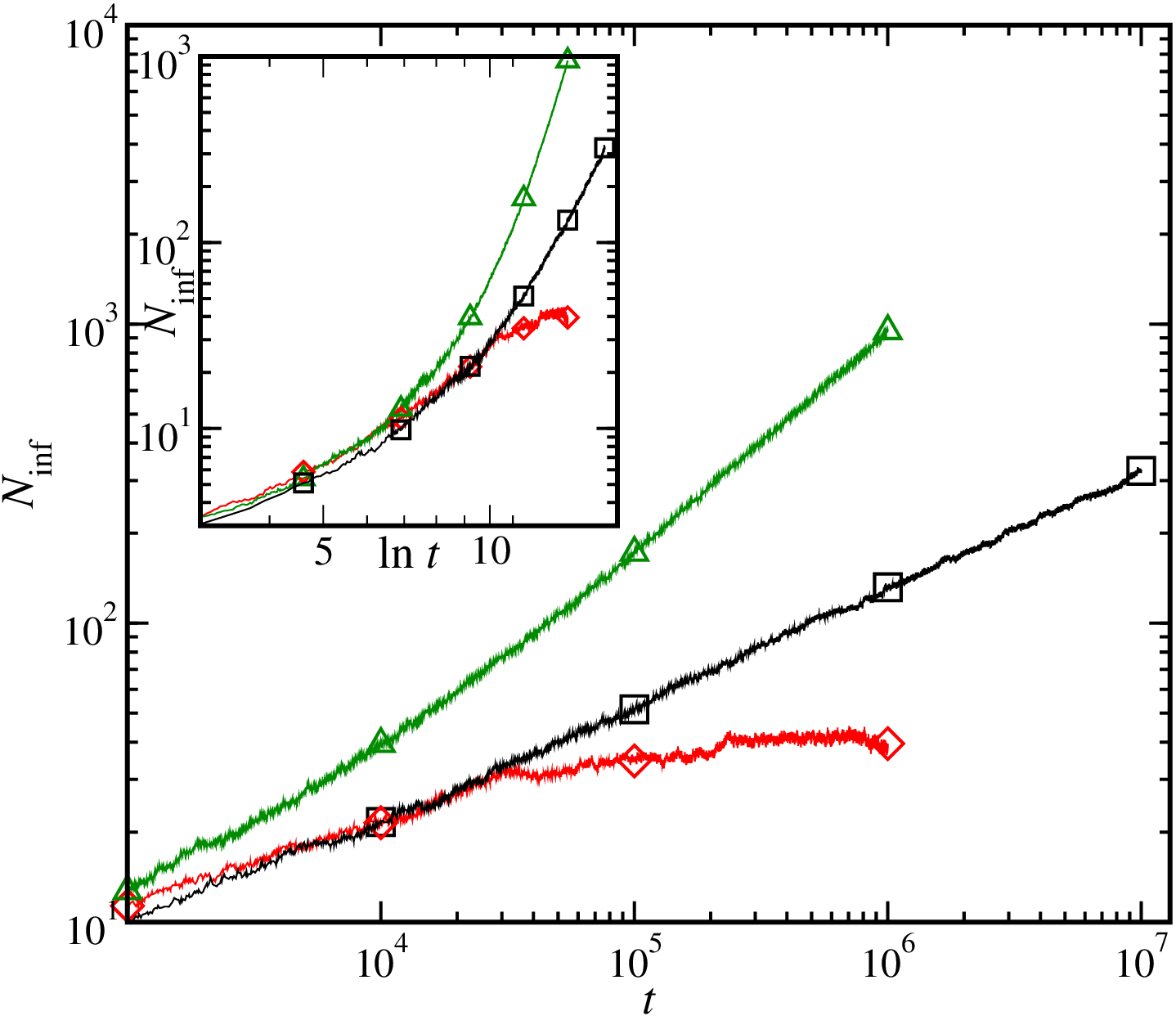}}
  \end{center}
  \caption{(Color online) Number of infected sites vs time obtained by MC
    simulations of the disordered CP with $q = 0.5$ and $\mu_A = 0.25$,
    averaged over $1500$ runs and at least $30$ disorder realizations. 
    The main figure is in double-log scale according to the conventional
    scaling while the inset demonstrates the same data in the 
    activated scaling picture \cite{vojta_05}. 
    The curves from top to bottom correspond to
    $0.3628$ ($\triangle$),
    $0.3680$ ($\Box$)
    and $0.3728$ ($\Diamond$).} 
    \label{fig:MC_bdcp}
\end{figure}

The results for the disorder case have been supported by MC simulations 
(see Fig.~\ref{fig:MC_bdcp}).  
Due to the long  relaxation times of the disordered CP
(cf. Refs.~\cite{noest_85, dickman_98, vojta_05}) we have focused only 
on one point in the rate space with $q = 0.5$ and $\mu_A=0.25$. 
The results of the simulations up to $10^7$ time steps are shown in
Fig.~\ref{fig:MC_bdcp} for three values of $\mu_B$ around criticality  
in double-log scales of $N_{\text{inf}}$ vs $t$
as well as $N_{\text{inf}}$ vs $\ln t$ (see the inset in Fig.~\ref{fig:MC_bdcp}) to
allow for both conventional and activated scaling \cite{vojta_05}. 
The MC critical value of $\mu_B \simeq 0.368$ (with the error being less than
0.005) obtained by conventional  
double-log scaling analysis is 
certainly very close to
the series expansion value
($\mu_B \simeq 0.369$ for $N=12$, see the triangle in Fig.~\ref{fig:bdcp_phase_diag}). 
The value of the dynamical exponent at this point is found to be $\eta \simeq 0.388$
which is in favour of the scenario suggesting scaling exponents
varying continuously with disorder.  

In conclusion, we have 
investigated the CP in heterogeneous and disordered 1$d$ systems in
the limit of weak disorder by means of the series expansions and MC
simulations. 
We have demonstrated that the CP in heterogeneous 1$d$ lattices stays in the DP
universality class. 
For disordered environment, our results suggest that
disorder continuously changes the scaling exponents. 
A simple analytical formula for the phase-separation line has been suggested
and proved (numerically) to be valid in the weak-disorder limit. 
Preliminary investigations of the CP in 2$d$ heterogeneous lattices also support 
the analytical predictions for the phase-separation line. 

\emph{Acknowledgements} - The authors would like to thank I.\ Jensen, B.\ Simons, J.\ Stilck, and R.\
Dickman for their helpful correspondence and remarks. The computations were
mainly performed on the Cambridge University Condor Grid and the
Cambridge-Cranfield High-Performance Computer
facility. CJN and SVF would like
to thank the EPSRC and the Cambridge European Trust for finacial
support.

\bibliographystyle{apsrev}

\begin{thebibliography}{24}
\expandafter\ifx\csname natexlab\endcsname\relax\def\natexlab#1{#1}\fi
\expandafter\ifx\csname bibnamefont\endcsname\relax
  \def\bibnamefont#1{#1}\fi
\expandafter\ifx\csname bibfnamefont\endcsname\relax
  \def\bibfnamefont#1{#1}\fi
\expandafter\ifx\csname citenamefont\endcsname\relax
  \def\citenamefont#1{#1}\fi
\expandafter\ifx\csname url\endcsname\relax
  \def\url#1{\texttt{#1}}\fi
\expandafter\ifx\csname urlprefix\endcsname\relax\def\urlprefix{URL }\fi
\providecommand{\bibinfo}[2]{#2}
\providecommand{\eprint}[2][]{\url{#2}}

\bibitem[{\citenamefont{Harris}(1974{\natexlab{a}})}]{harris_74}
\bibinfo{author}{\bibfnamefont{T.}~\bibnamefont{Harris}},
  \bibinfo{journal}{Annals of Probability} \textbf{\bibinfo{volume}{2}},
  \bibinfo{pages}{969} (\bibinfo{year}{1974}{\natexlab{a}}).

\bibitem[{\citenamefont{Hinrichsen}(2000)}]{hinrichsen_00}
\bibinfo{author}{\bibfnamefont{H.}~\bibnamefont{Hinrichsen}},
  \bibinfo{journal}{Advances in Physics} \textbf{\bibinfo{volume}{49}},
  \bibinfo{pages}{815} (\bibinfo{year}{2000}).

\bibitem[{\citenamefont{Harris}(1974{\natexlab{b}})}]{harris_74_2}
\bibinfo{author}{\bibfnamefont{A.~B.} \bibnamefont{Harris}},
  \bibinfo{journal}{J. Phys. C} \textbf{\bibinfo{volume}{7}},
  \bibinfo{pages}{1671} (\bibinfo{year}{1974}{\natexlab{b}}).

\bibitem[{\citenamefont{Chayes et~al.}(1986)\citenamefont{Chayes, Chayes,
  Fisher, and Spencer}}]{chayes_86}
\bibinfo{author}{\bibfnamefont{J.}~\bibnamefont{Chayes et~al.}},
  \bibinfo{journal}{Phys. Rev. Lett.} \textbf{\bibinfo{volume}{57}},
  \bibinfo{pages}{2999} (\bibinfo{year}{1986}).

\bibitem[{\citenamefont{Noest}(1986)}]{noest_85}
\bibinfo{author}{\bibfnamefont{A.}~\bibnamefont{Noest}},
  \bibinfo{journal}{Phys. Rev. Lett.} \textbf{\bibinfo{volume}{57}},
  \bibinfo{pages}{90} (\bibinfo{year}{1986}).

\bibitem[{\citenamefont{Dickman and Moreira}(1996)}]{dickman_96}
\bibinfo{author}{\bibfnamefont{A.}~\bibnamefont{Moreira}} \bibnamefont{and}
  \bibinfo{author}{\bibfnamefont{R.}~\bibnamefont{Dickman}},
  \bibinfo{journal}{Phys. Rev. E} \textbf{\bibinfo{volume}{54}},
  \bibinfo{pages}{R3090}(\bibinfo{year}{1996}).

\bibitem[{\citenamefont{Dickman and Moreira}(1998)}]{dickman_98}
\bibinfo{author}{\bibfnamefont{R.}~\bibnamefont{Dickman}} \bibnamefont{and}
  \bibinfo{author}{\bibfnamefont{A.}~\bibnamefont{Moreira}},
  \bibinfo{journal}{Phys. Rev. E} \textbf{\bibinfo{volume}{57}},
  \bibinfo{pages}{1263} (\bibinfo{year}{1998}).

\bibitem[{\citenamefont{Janssen}(1997)}]{janssen_97}
\bibinfo{author}{\bibfnamefont{H.~K.} \bibnamefont{Janssen}},
  \bibinfo{journal}{Phys. Rev. E} \textbf{\bibinfo{volume}{55}},
  \bibinfo{pages}{6253} (\bibinfo{year}{1997}).

\bibitem[{\citenamefont{Hooyberghs et~al.}(2003)\citenamefont{Hooyberghs,
  Igl\'oi, and Vanderzande}}]{hooyberghs_03}
\bibinfo{author}{\bibfnamefont{J.}~\bibnamefont{Hooyberghs et~al.}},
  \bibinfo{journal}{Phys. Rev. Lett.} \textbf{\bibinfo{volume}{90}},
  \bibinfo{pages}{100601} (\bibinfo{year}{2003}).

\bibitem[{\citenamefont{Hooyberghs et~al.}(2004)\citenamefont{Hooyberghs,
  Igl\'oi, and Vanderzande}}]{hooyberghs_04}
\bibinfo{author}{\bibfnamefont{J.}~\bibnamefont{Hooyberghs et~al.}},
  \bibinfo{journal}{Phys. Rev. E} \textbf{\bibinfo{volume}{69}},
  \bibinfo{pages}{66140} (\bibinfo{year}{2004}).

\bibitem[{\citenamefont{Vojta and Dickison}(2005)}]{vojta_05}
\bibinfo{author}{\bibfnamefont{T.}~\bibnamefont{Vojta}} \bibnamefont{and}
  \bibinfo{author}{\bibfnamefont{M.}~\bibnamefont{Dickison}},
  \bibinfo{journal}{Phys. Rev. E} \textbf{\bibinfo{volume}{72}},
  \bibinfo{pages}{36126} (\bibinfo{year}{2005}).

\bibitem[{\citenamefont{Dickman and Jensen}(1991)}]{dickman_91}
\bibinfo{author}{\bibfnamefont{R.}~\bibnamefont{Dickman}} \bibnamefont{and}
  \bibinfo{author}{\bibfnamefont{I.}~\bibnamefont{Jensen}},
  \bibinfo{journal}{Phys. Rev. Lett.} \textbf{\bibinfo{volume}{67}},
  \bibinfo{pages}{2391} (\bibinfo{year}{1991}).

\bibitem[{\citenamefont{Jensen and Dickman}(1993)}]{jensen_93}
\bibinfo{author}{\bibfnamefont{I.}~\bibnamefont{Jensen}} \bibnamefont{and}
  \bibinfo{author}{\bibfnamefont{R.}~\bibnamefont{Dickman}},
  \bibinfo{journal}{J. Stat. Phys.} \textbf{\bibinfo{volume}{71}},
  \bibinfo{pages}{89} (\bibinfo{year}{1993}).

\bibitem[{\citenamefont{Liggett}(1985)}]{liggett_85}
\bibinfo{author}{\bibfnamefont{T.}~\bibnamefont{Liggett}},
  \emph{\bibinfo{title}{Interacting Particle Systems}} (\bibinfo{year}{1985}),
  \bibinfo{edition}{1st} ed.

\bibitem[{\citenamefont{Jensen}(1996)}]{jensen_96}
\bibinfo{author}{\bibfnamefont{I.}~\bibnamefont{Jensen}}, \bibinfo{journal}{J.
  Phys. A} \textbf{\bibinfo{volume}{29}}, \bibinfo{pages}{7013}
  (\bibinfo{year}{1996}).

\bibitem[{\citenamefont{Bramson et~al.}(1991)\citenamefont{Bramson, Durrett,
  and Schonmann}}]{bramson_91}
\bibinfo{author}{\bibfnamefont{M.}~\bibnamefont{Bramson et~al.}},
  \bibinfo{journal}{Ann. Probab.} \textbf{\bibinfo{volume}{19}},
  \bibinfo{pages}{960} (\bibinfo{year}{1991}).

\bibitem[{\citenamefont{Fisher}(1995)}]{fisher_95}
\bibinfo{author}{\bibfnamefont{D.~S.} \bibnamefont{Fisher}},
  \bibinfo{journal}{Phys. Rev. B} \textbf{\bibinfo{volume}{51}},
  \bibinfo{pages}{6411} (\bibinfo{year}{1995}).

\bibitem[{\citenamefont{Doi}(1976)}]{doi_76}
\bibinfo{author}{\bibfnamefont{M.}~\bibnamefont{Doi}}, \bibinfo{journal}{J.
  Phys. A} \textbf{\bibinfo{volume}{9}}, \bibinfo{pages}{1479}
  (\bibinfo{year}{1976}).

\bibitem[{\citenamefont{Grassberger and De~La~Torre}(1979)}]{grassberger_79}
\bibinfo{author}{\bibfnamefont{P.}~\bibnamefont{Grassberger}} \bibnamefont{and}
  \bibinfo{author}{\bibfnamefont{A.}~\bibnamefont{De~La~Torre}},
  \bibinfo{journal}{Ann. Phys. (N.Y.)} \textbf{\bibinfo{volume}{122}},
  \bibinfo{pages}{373} (\bibinfo{year}{1979}).

\bibitem[{\citenamefont{Peliti}(1985)}]{peliti_85}
\bibinfo{author}{\bibfnamefont{L.}~\bibnamefont{Peliti}}, \bibinfo{journal}{J.
  Physique} \textbf{\bibinfo{volume}{46}}, \bibinfo{pages}{1469}
  (\bibinfo{year}{1985}).

\bibitem[{\citenamefont{Dantas and Stilck}(2005)}]{dantas_05}
\bibinfo{author}{\bibfnamefont{W.~G.} \bibnamefont{Dantas}} \bibnamefont{and}
  \bibinfo{author}{\bibfnamefont{J.~F.} \bibnamefont{Stilck}},
  \bibinfo{journal}{J. Phys. A: Math. Gen.} \textbf{\bibinfo{volume}{38}},
  \bibinfo{pages}{5841} (\bibinfo{year}{2005}).

\bibitem[{\citenamefont{Guillaume}(1997)}]{guillaume_97}
\bibinfo{author}{\bibfnamefont{P.}~\bibnamefont{Guillaume}},
  \bibinfo{journal}{J. Comp. Appl. Math.} \textbf{\bibinfo{volume}{82}},
  \bibinfo{pages}{149} (\bibinfo{year}{1997}).

\bibitem[{\citenamefont{Guillaume and Huard}(2000)}]{guillaume_00}
\bibinfo{author}{\bibfnamefont{P.}~\bibnamefont{Guillaume}} \bibnamefont{and}
  \bibinfo{author}{\bibfnamefont{A.}~\bibnamefont{Huard}}, \bibinfo{journal}{J.
  Comp. Appl. Math.} \textbf{\bibinfo{volume}{121}}, \bibinfo{pages}{197}
  (\bibinfo{year}{2000}).

\bibitem[{\citenamefont{Marro and Dickman}(1999)}]{marro_99}
\bibinfo{author}{\bibfnamefont{J.}~\bibnamefont{Marro}} \bibnamefont{and}
  \bibinfo{author}{\bibfnamefont{R.}~\bibnamefont{Dickman}},
  \emph{\bibinfo{title}{Nonequilibrium Phase Transitions in Lattice Models}}
  (\bibinfo{publisher}{CUP}, \bibinfo{year}{1999}, \bibinfo{edition}{1st} ed.).

\end{thebibliography}

\end{document}